\documentclass[twocolumn,showpacs,preprintnumbers,amsmath,amssymb]{revtex4}

\usepackage{graphicx}
\usepackage{dcolumn}
\usepackage{bm}

\begin{document}

\title{Self-Organizing Traffic Lights}
\author{Carlos Gershenson}
 \affiliation{%
Centrum Leo Apostel, Vrije Universiteit Brussel\\
Krijgskundestraat 33 B-1160 Brussel, Belgium
}%
 \email{cgershen@vub.ac.be}
 \homepage{http://homepages.vub.ac.be/~cgershen}

\date{\today}

\begin{abstract}
Steering traffic in cities is a very complex task, since improving
efficiency involves the coordination of many actors. Traditional approaches
attempt to optimize traffic lights for a particular density and
configuration of traffic. The disadvantage of this lies in the fact that
traffic densities and configurations change constantly. Traffic seems to be
an adaptation problem rather than an optimization problem. We propose a
simple and feasible alternative, in which traffic lights \emph{self-organize}
to improve traffic flow. We use a multi-agent simulation to study three
self-organizing methods, which are able to outperform traditional rigid and
adaptive methods. Using simple rules and no direct communication, traffic
lights are able to self-organize and adapt to changing traffic conditions,
reducing waiting times, number of stopped cars, and increasing average
speeds.
\end{abstract}


\maketitle

\section{Introduction}

Anyone living in a populated area suffers from traffic congestion. Traffic
is time, energy, and patience consuming. This has motivated people to
regulate traffic flow in order to reduce the congestion. The idea is simple:
if vehicles are allowed to go in any direction, there is a high probability
that one will obstruct another. To avoid this, rules have been introduced to 
\emph{mediate} \cite{Heylighen2004} between the conflicting vehicles, by
restricting or bounding their behaviour. People have agreed on which side of
the street they will drive (left or right); traffic lanes prevent cars from
taking more space than necessary; traffic signals and codes prompt an
appropriate behaviour; and traffic lights regulate the crossing of
intersections.

There is no solution to the traffic congestion problem when the car density
saturates the streets, but there are many ways in which the car flow can be 
\emph{constrained} in order to improve traffic. Traffic lights are not the
only component to take into account, but they are an important factor. We
can say that a traffic light system will be more efficient if, for a given
car density, it increases the average speeds of vehicles. This is reflected
in less time that cars will wait behind red lights.

For decades, people have been using mathematical and computational methods
that find appropriate periods and phases (i.e. cycles) of traffic lights, so
that the variables considered will be \emph{optimized}. This is good because
certain synchronization is better than having no correlation of phases.
However, many methods applied today do not consider the current state of the
traffic. If cars are too slow for the expected average speed, this might
result in the loss of the phases dictated by the traffic lights. If they go
too fast, they will have to wait until the green light phase reaches every
intersection. The optimizing methods are \emph{blind} to "abnormal"
situations, such as many vehicles arriving or leaving a certain place at the
same time, e.g. a stadium, a financial district, a university. In most
cases, traffic agents need to override the traffic lights and personally
regulate the traffic. Nevertheless, traffic modelling has improved greatly
our understanding of this complex phenomenon, especially during the last
decade \cite%
{PrigogineHerman1971,Traffic95,Traffic97,Traffic99,Helbing1997,HelbingHuberman1998}%
, suggesting different improvements to the traffic infrastructure.

We believe that traffic light control is not so much an optimization
problem, but rather an \emph{adaptation} problem, since traffic flows and
densities change constantly. Optimization gives the best possible solution
for a given configuration. But since the configuration is changing
constantly in real traffic, it seems that we would do better with an
adaptive mechanism than with a mechanism that is optimal some times, and
some times creates havoc. Indeed, modern "intelligent" advanced traffic
management systems (ATMS) use learning methods to adapt phases of traffic
lights, normally using a central computer\footnote{%
A drawback of ATMS is their high cost and complexity that requires
maintenance by specialists. There is yet no standard, and usually companies
should be hired to develop particular solutions for different cities.}\cite%
{FHA1998,SCOOT1981}. Another reason for preferring an adaptive method is
that optimization can be computationally expensive. Trying to find all
possible optimal solutions of a city is not feasible, since the
configuration space is too huge, uncertain, and it changes constantly.

In this paper, we present three simple \emph{traffic-responsive} methods for
traffic light control that are adaptive by \emph{self-organization}, and
compare them with two \emph{fixed-cycle} non-adaptive methods and another
traffic-responsive method. We use multi-agent computer simulations to do
this. In the next section, we make a brief and practical introduction to the
concept of self-organization. Then we present the simulation and the control
methods we compared. We show first results in Section \ref{S-FirstResults}.
We present improvements we did to our simulation to make it more realistic
in Section \ref{S-Improvements}. The results of further experiments are
shown in Section \ref{S-SecondResults}. We discuss the results and
implications in Section \ref{S-Discussion} and conclude in Section \ref%
{S-Conclusions}.

\section{Self-organization}

The term \emph{self-organization} has been used in different areas with
different meanings, such as cybernetics \cite{vonFoerster1960,Ashby1962},
thermodynamics \cite{NicolisPrigogine1977}, mathematics \cite{Lendaris1964},
computing \cite{HeylighenGershenson2003},information theory \cite%
{Shalizi2001}, synergetics \cite{Haken1981}, and others (for a general
overview, see \cite{Heylighen2003sos}). However, the use of the term is
subtle, since any dynamical system can be said to be self-organizing or not,
depending partly on the observer \cite{GershensonHeylighen2003a,Ashby1962}.

Without entering into a philosophical debate on the theoretical aspects of
self-organization, a practical definition will suffice for our present work.
For us, a system described as self-organizing is one in which elements \emph{%
interact} in order to achieve a global function or behaviour. This function
or behaviour is not imposed by a single or few elements, nor determined
hierarchically. It is achieved dynamically as the elements interact with one
another. These interactions produce feedbacks that regulate the system.

Many distributed adaptive traffic light systems can be considered as
self-organizing, e.g. \cite{SCATS1979,FaietaHuberman1993}. Nevertheless, the
methods presented in this paper distinguish themselves because there is no
communication between traffic lights, only local rules (an analysis of their
indirect interactions is given in Section \ref{S-Discussion}). Still, they
are able to achieve global coordination of traffic.

We believe that this approach is useful for systems such as traffic lights,
since the "solution" of the problem is not known beforehand, but strived for
dynamically by the elements of the system. In this way, systems can adapt
quickly to unforeseen changes as elements interact locally. It should be
noted that self-organizing approaches are being used in other areas of
traffic control e.g. \cite{WischhofEtAl2003}.

The present work is very abstract. The models presented were not developed
to be directly applied on real scenarios (more realistic simulations and
pilot studies would be required), but to explore and understand principles
of self-organization in traffic light control. The next section describes
the simulation where we test various models

\section{The Simulation}

Several traffic simulations use cellular automata to model traffic
effectively \cite{BML1992,NaSch1992,ChSch1999,FaietaHuberman1993}, since it
is computationally cheaper. However, the increase of computing power in the
last few years has allowed the development of multi-agent simulations to
create more realistic traffic simulations \cite%
{Nagel2004,WieringEtAl2004,Miramontes2004,RoozemondRogier2000}.

\begin{figure}[tbp]
\begin{center}
\includegraphics[
natheight=4.4304in, natwidth=4.2921in, height=5.7024cm, width=5.5201cm]
{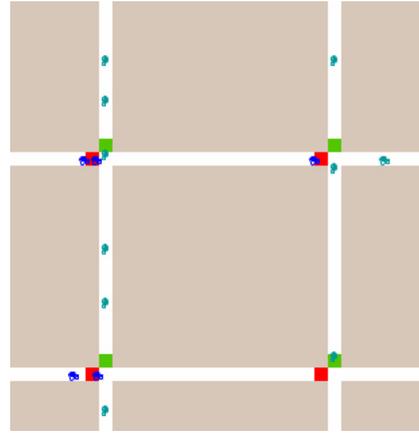}
\end{center}
\caption{Screenshot of part of traffic grid. Green lights southbound, red
light eastbound. (Color online).}
\label{screenshot}
\end{figure}

We developed a simulation in NetLogo \cite{Wilensky1999}, a multi-agent
modelling environment. We extended the "Gridlock" model \cite%
{WilenskyStroup2002} which is included in the NetLogo distribution. It
consists of an abstract traffic grid with intersections between cyclic
single-lane arteries of two types: vertical or horizontal. In the first
series of experiments, similar to the scenario of \cite{BrockfeldEtAl2001},\
cars only flow in a straight line, either eastbound or southbound. Each
crossroad has traffic lights which allow traffic flow in only one of the
arteries which intersect it with a green light. Yellow or red lights stop
the traffic. The light sequence for a given artery is
green-yellow-red-green. Cars simply try to go at a maximum speed of 1
"patch" per timestep, but stop when a car or a red or yellow light is in
front of them. Time is discrete, but not space. A "patch" is a square of the
environment the size of a car. A screenshot of the environment can be seen
in Figure \ref{screenshot}. The reader is invited to test the simulation
(source code included), with the aid of a Java-enabled Internet browser, at
the URL http://homepages.vub.ac.be/\symbol{126}cgershen/sos/SOTL/SOTL.html .

The user can change different parameters, such as the number of arteries or
number of cars. Different statistics are shown: the number of stopped cars,
the average speed of cars, and the average waiting times of cars.

\section{The Control Methods}

\subsection{"Marching" control}

This is a very simple method. All traffic lights "march in step": all green
lights are either southbound or eastbound, synchronized in time.
Intersections have a phase $\varphi _{i}$, which counts time steps. $\varphi
_{i}$ is reset to zero when the phase reaches a period value $p$. When $%
\varphi _{i}==0$, red lights turn green, and yellow lights turn red. Green
lights turn yellow one time step earlier, i.e. when $\varphi ==p-1$. A full
cycle of an intersection consists of $2p$ time steps. "Marching"
intersections are such that $\varphi _{i}==\varphi _{j},\forall i,j.$

\subsection{"Optim" control}

This method is implemented trying to set phases $\varphi _{i}$ of traffic
lights in such a way that, as soon as a red light turns green, a car stopped
by this would find the following traffic lights green. In other words, we
obtain a fixed solution so that \emph{green waves} flow to the southeast.

The simulation environment has a radius of $r$ square patches, so that these
can be identified with coordinates $(x_{i},y_{i}),$ $x_{i},y_{i}\in \lbrack
-r,r]$. Therefore, each artery consists of $2r+1$ number of patches (In the
presented results, $r=80$, but this can be easily varied in the simulation).
In order to synchronize all the intersections (which occupy one patch each),
red lights should turn green and yellow lights should turn red when

\begin{equation}
\varphi _{i}==round(\frac{2r+x_{i}-y_{i}}{4})
\end{equation}

and green lights should turn to yellow the previous time step. The period
should be $p=r+3$. The three is added as an extra margin for the reaction
and acceleration times of cars (found to be best, for low densities, by
trial and error).

A disadvantage of the \emph{optim} control is that the average speed
decreases as the traffic density increases, so cars don't manage to keep up
the speed of the "green waves". A different solution could be obtained, for
lower average speeds, but then the green waves would be too slow for low
traffic densities\footnote{%
Some real traffic light systems have different "optimal" solutions (i.e.
different $p$ and $\varphi _{i}$ values) for different times of the day \cite%
{FHA1998}.}.

These two first methods are \emph{non-adaptive}, in the sense that their
behaviour is dictated beforehand, and they do not consider the actual state
of the traffic.

\subsection{"Sotl-request" control}

All three self-organizing control methods use a similar principle: traffic
lights keep a count $\kappa _{i}$ of the number of cars times time steps ($%
c\ast ts$) approaching \emph{only} the red light, independently of the
status or speed of the cars (i.e. moving or stopped). When $\kappa _{i}$
reaches a threshold $\theta $, the opposing green light turns yellow, and
the following time step it turns red with $\kappa _{i}=0$ , while the red
light which counted turns green. In this way, if there are more cars
approaching or waiting behind a red light, this will turn into green faster
than if there are only few cars. This simple mechanism achieves
self-organization in the following way: if there are single or few cars,
these will be stopped for more time behind red lights. This gives time for
other cars to join them. As more cars join the group, cars will wait less
time behind red lights. With a sufficient number of cars, the red lights
will turn green even before they reach the intersection, generating "green
corridors". Having "platoons" or "convoys" of cars moving together improves
traffic flow, compared to a homogeneous distribution of cars, since there
are large empty areas between platoons, which can be used by crossing
platoons with few interferences.

The \emph{sotl-request} method has no phase or internal clock. Traffic
lights change only when the above conditions are met. If there are no cars
approaching a red light, the complementary one can stay green. However,
depending on the value of $\theta $, high traffic densities can trigger
light switching too fast, obstructing traffic flow.

It is worth mentioning that this method was discovered "by accident". This
was due to an unintended error in the programming while testing a different
control method.

\subsection{"Sotl-phase" control}

The \emph{sotl-phase} method differs from \emph{sotl-request} adding the
following constraint: A traffic light will not be changed if the number of
time steps is less than a minimum phase, i.e. $\varphi _{i}<\varphi _{\min }$%
. Once $\varphi _{i}\geq \varphi _{\min }$, the lights will change if/when $%
\kappa _{i}$ $\geq $ $\theta $. This prevents the fast switching of lights%
\footnote{%
A similar method has been used successfully in the United Kingdom for some
time, but for isolated intersections \cite{VincentYoung1986}.}.

\subsection{"Sotl-platoon" control}

The \emph{sotl-platoon} method adds two further restrictions to \emph{%
sotl-phase} to regulate the size of platoons. Before changing a red light to
green, it checks if a platoon is not crossing through, in order not to break
it. More precisely, a red light is not changed to green if on the crossing
street there is at least one car approaching at $\omega $ patches from the
intersection. This keeps platoons together. For high densities, this
restriction alone would cause havoc, since large platoons would block the
traffic flow of intersecting streets. To avoid this, we introduce a second
restriction. Restriction one is not taken into account if there are more
than $\mu $ cars approaching the intersection. Like this, long platoons can
be broken, and the restriction only comes into place if a platoon will soon
be through an intersection.

Curiously, this method was the result of misinterpreting a suggestion by
Bart De Vylder.

We say that these three adaptive methods are self-organizing because the
global performance is given by the local rules followed by each traffic
light: they are unaware of the state of other intersections and still manage
to achieve global coordination.

The \emph{sotl} methods use a similar idea to the one used by \cite[and
references within]{PorcheLafortune1998}, but with a much simpler
implementation. There is no costly prediction of arrivals at intersections,
and no need to establish communication between traffic lights to achieve
coordination. They do not have fixed cycles.

\subsection{"Cut-off" control}

We wanted to compare our self-organizing methods with a traditional traffic
responsive method, that has proven to be better than static methods at
single intersections \cite{FouladvandEtAl2004a}. The idea of the \emph{%
cut-off} method is simple: a traffic light will remain green until a queue
of \emph{stopped} waiting cars reaches a length of $\lambda $ cars. At this
moment, the green light turns yellow, and at the next time step, red, while
the opposing light turns green.

Recall that \emph{sotl} methods keep a count of approaching cars,
independently of their speed. Therefore, cars do not need to stop in order
to change a traffic light.

\subsection{"No-corr" control}

To have an idea of the benefit of the different control methods, we also
compared them with a non-correlated scheme: each traffic light is assigned a
phase $\varphi _{i}$ at random, and remains fixed during a simulation run.
There is no correlation between different intersections.

\section{First Results\label{S-FirstResults}}

\begin{figure*}[tbp]
\begin{center}
\includegraphics[
natheight= 10.2359in, natwidth=6.6599in, 
height=9.0000in, width=5.8558in]
{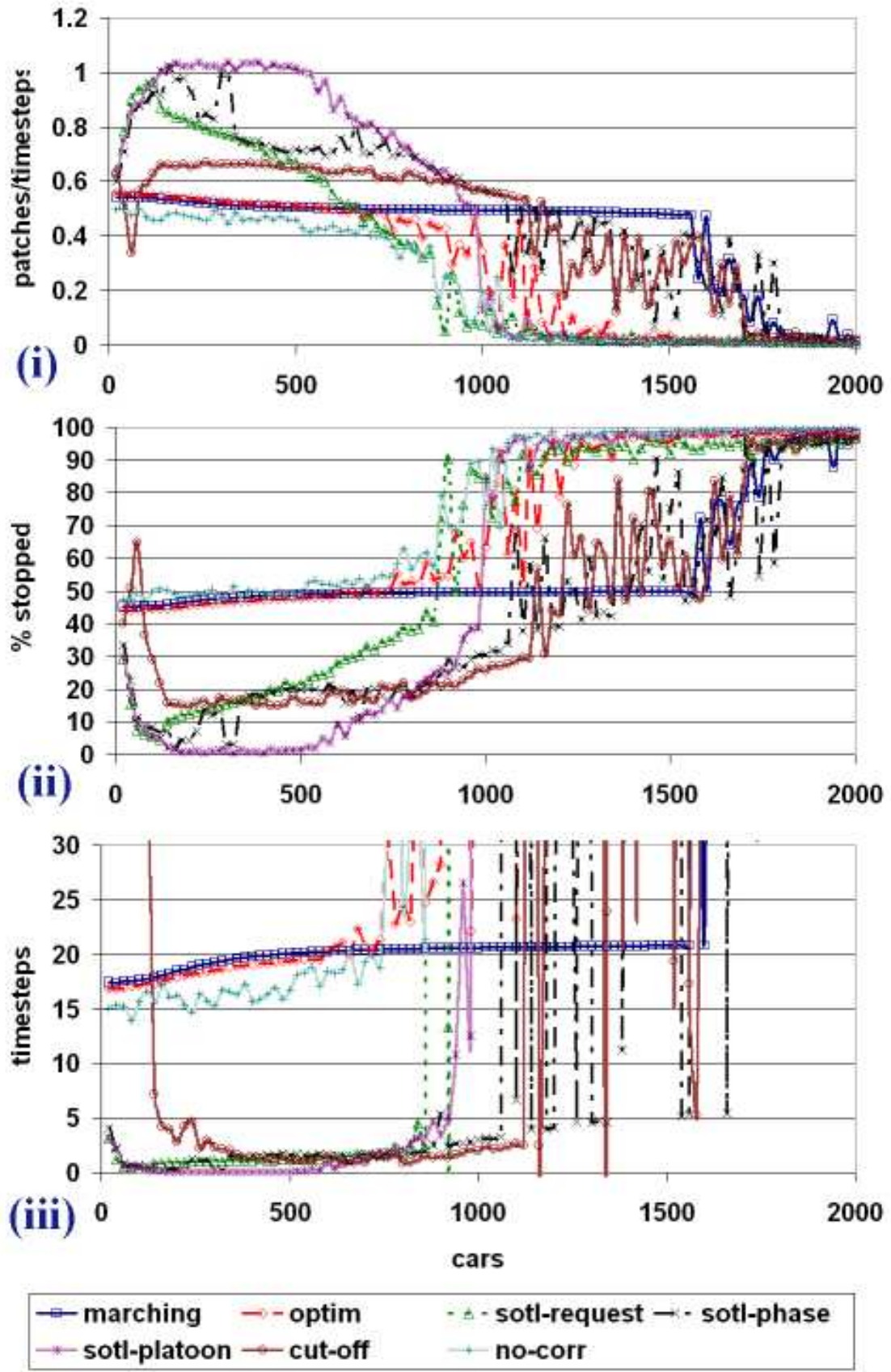}
\end{center}
\caption{(i) Average speeds of cars. (ii) Percentage of stopped cars. (iii)
Average waiting times. Very high waiting times (out of graph) indicate
deadlocks. (Color online).}
\label{2dresults}
\end{figure*}

We performed simulations in order to obtain average statistics of the
performance of the different control methods. These were namely speed%
\footnote{%
The cruise speed is 1 patch/time step, i.e. the speed at which cars go
without obstructions}, percentage of stopped cars, and waiting time. The
results shown in Figure \ref{2dresults} were obtained in a grid of 10x10
arteries of $r=80$ (therefore 3120 available patches), with $p=83$, $\theta
=41$, $\varphi _{\min }=20$, $\omega =4$, $\mu =3$, $\lambda =3$. We did for
each method one run varying the number of cars from twenty to two thousand,
in steps of twenty (one hundred and one runs in total), with the same
parameters.

We can see that the \emph{marching} method is not very efficient for low
traffic densities. Since half of the arteries (all eastbound or all
southbound) have red lights, this causes almost half of the cars to be
stopped at any time, reducing the average speed of cars. On the other hand,
its performance degrades slowly as the traffic densities reach certain
levels, and performs the best for very high densities. This is because it
keeps a strict division of space occupied by cars, and interferences are
less probable.

For low densities, the \emph{optim} method performs acceptably. However, for
high densities cars can enter a deadlock much faster than with other
methods. This is because cars waiting behind other cars in red lights do not
reach green waves, reducing their speed and the speed of the cars which go
behind them. Also, even when there will be some cars that do not stop,
flowing through green waves, there will be an equivalent number of cars
waiting to enter a green wave, losing the time gained by cars in green
waves. Therefore, the performance cannot be much better than \emph{marching}.

\emph{Sotl-request} gives the best performance for low traffic densities
because platoons can quickly change red lights into green, in most cases
before actually reaching the intersections. Since the traffic density is
low, this does not obstruct many cars approaching the intersection in the
corresponding artery. However, for high densities this method is extremely
inefficient, since there is a constant switching of lights due to the fact
that $\theta $ is reached very fast. This reduces the speed of cars, since
they stop on yellow lights, but also breaks platoons, so that the few cars
that pass will have a higher probability of waiting more time at the next
intersection.

\begin{figure}[tbp]
\begin{center}
\includegraphics[
natheight=7.8611in, natwidth=3.8337in, height=5.5305in, width=2.7112in]
{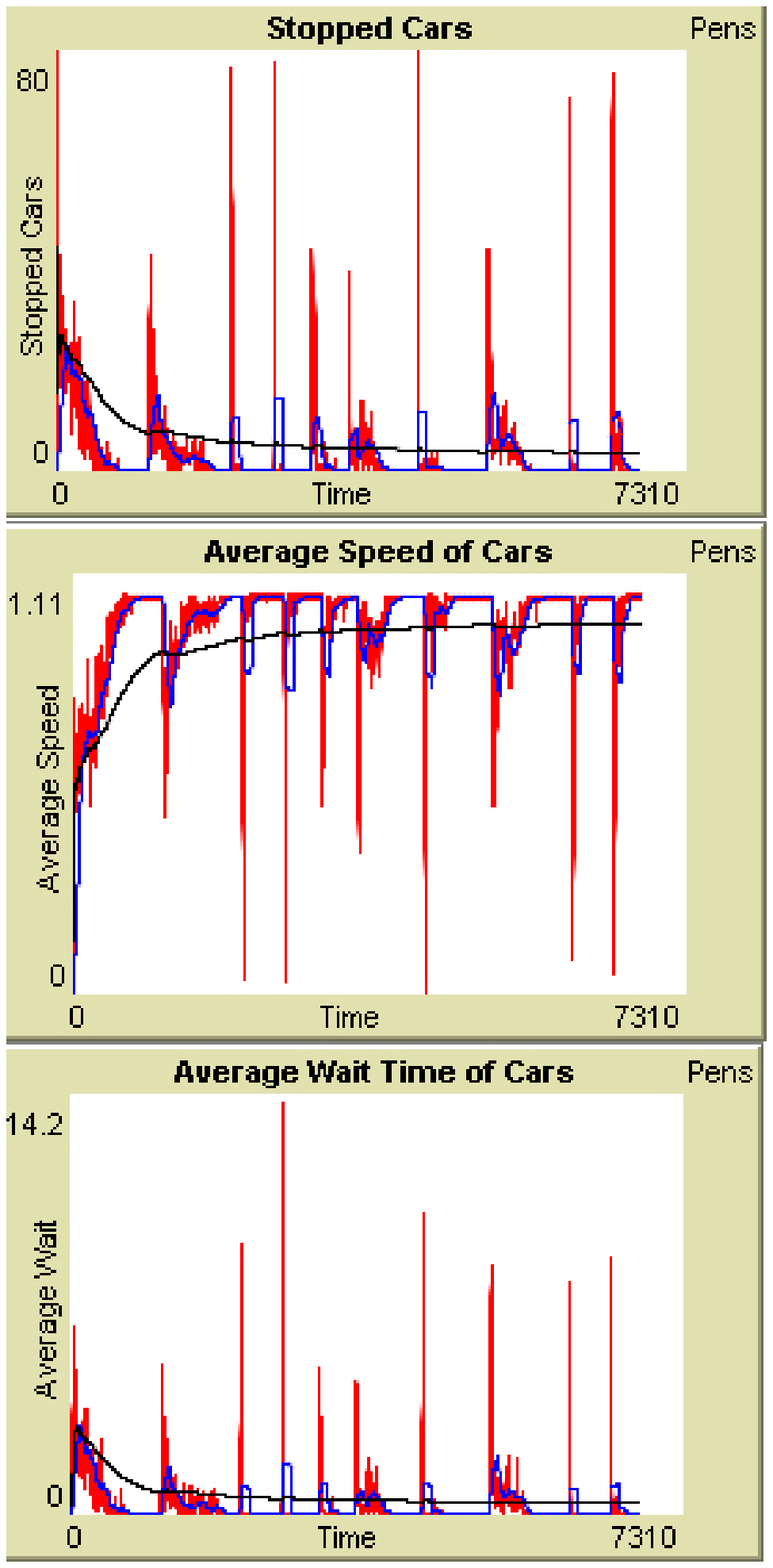}
\end{center}
\caption{Resets of traffic lights as \emph{sotl-phase} achieves full
synchronization (80 cars in 5x5 grid, $r=40$). (Color online).}
\label{resetTL}
\end{figure}

\emph{Sotl-phase} does not perform as good as \emph{sotl-request} for low
densities because in many cases cars need to wait in front of red lights as $%
\kappa _{i}$ reaches $\varphi _{\min }$, with no cars coming in the
corresponding artery. The performance of \emph{sotl} methods could be
improved for low densities by reducing $\theta $, since small platoons might
need to wait too long at red lights. As the traffic density reaches a medium
scale, platoons effectively exploit their size to accelerate their
intersection crossing. With the considered parameters, in the region around
160 cars, and again at around 320, sotl-phase can achieve \emph{full
synchronization} in space, in the sense that no platoon has to stop, so all
cars can go at a maximum speed. This is not a realistic situation, because
synchronization is achieved due to the toroidal topology of the simulation
environment. Still, it is interesting to understand the process by which the
full synchrony is reached. Platoons are formed, as described in the previous
section, of observed sizes $3\leq cars\leq 15$. One or two platoons flow per
street. Remember that platoons can change red lights to green before they
reach an intersection, if $\kappa _{i}\geq \varphi _{\min }$. If a platoon
moving in an artery is obstructed, this will be because still $\kappa
_{i}<\varphi _{\min }$, and because a platoon is crossing, or crossed the
intersection recently in the complementary artery. The waiting of the
platoon will change its phase compared to other flowing platoons. However,
if no platoon crossed recently, a platoon will keep its phase relative to
other platoons. This induces platoons not to interfere with each other,
until all of them go at maximum speed. We can see that this condition is
robust by resetting the traffic light periods and $\kappa _{i}$'s. Each
reset can be seen in the spikes of the graphs shown in Figure \ref{resetTL}.
Nevertheless, the precise time in which the full synchronization is reached
can vary. For some initial conditions, full synchronization is not achieved,
but it is approached nevertheless.

The phenomenon of full synchronization shows us how self-organizing traffic
lights form platoons, which in turn modulate traffic lights. This feedback
is such that it maximizes average speeds and minimizes waiting times and
stopped cars in a robust way. The self-organizing traffic lights are
efficient without knowing beforehand the locations or densities of cars.

When there is a very high traffic density, \emph{optim} and \emph{%
sotl-request} reach deadlocks frequently, where all traffic is stopped. 
\emph{Sotl-phase} behaves similar to \emph{marching}, since traffic lights
change as soon as $\kappa _{i}\geq \varphi _{\min }$, because in most cases $%
\kappa _{i}\geq \theta $ by then. This also reduces the sizes of platoons,
which if very long can generate deadlocks. However, when the traffic density
is too high, deadlocks will be inevitable, though \emph{marching} generates
less deadlocks than \emph{sotl-phase}. This is because with the \emph{%
marching} method whole arteries are either stopped or advancing. This
reduces the probability of having a green light where cars cannot cross
(e.g. due to a red light ahead, and a line of cars waiting to cross it),
which would block the crossing artery at the next phase\footnote{%
Deadlocks could be avoided by restricting all cars to cross intersections
unless there is free space after it. However, it is unrealistic to expect
human drivers to behave in this way.}.

\emph{Sotl-platoon} manages to keep platoons together, achieving full
synchronization commonly for a wide density range, more effectively than 
\emph{sotl-phase}. This is because the restrictions of this method prevent
platoons from leaving few cars behind, with a small time cost for waiting
vehicles. Still, this cost is much lower than breaking a platoon and waiting
for separated vehicles to join back again. A platoon is divided only if $\mu
=3$, and a platoon of size three will manage to switch traffic lights
without stopping for the simulation parameters used. However, for high
traffic densities platoons aggregate too much, making traffic jams more
probable. The \emph{sotl-platoon} method fails when a platoon waiting to
cross a street is long enough to reach the previous intersection, but not
long enough to cut its tail. This will prevent waiting cars from advancing,
until more cars join the long platoon. This failure could probably be
avoided introducing further restrictions in the method, but here we would
like to study only very simple methods.

The platoon size in \emph{sotl} strategies depends on the tolerance $\theta $
and the distance between crossings, since longer distances give more time to 
$\kappa _{i}$ to reach $\theta $. An alternative would be to count cars at a
specified distance, independently of the distance between crossings, so that
the method could be also useful when traffic lights are very close together,
or far away. This should also be considered in a non-homogeneous grid.

\emph{Cut-off} performs better than the static methods, as it responds to
the current traffic state (except for very low densities, when cars in
streets may never reach the cut-off length $\lambda $). However, it is not
as efficient as \emph{sotl} methods, since cars need to stop before being
able to switch a red light to green. Still, for high densities its
performance is comparable to that of \emph{sotl-phase}, performing better
than the other two \emph{sotl} methods.

With \emph{no-corr}, we can observe that all the methods have an improvement
over random phase assignation. Nevertheless, the difference between \emph{%
no-corr} and static methods is less than the one between static and adaptive
methods. This suggests that, for low traffic densities, adaptation is more
important than "blind" correlation. For high traffic densities, the opposite
seems to be the case. Still, adaptive methods have correlation inbuilt.

We performed tests with "faulty" i.e. non-correlated intersections. All
methods are robust to failure of synchronization of individual traffic
lights, and the global performance degrades gracefully as more traffic
lights become faulty.

\section{Improvements to Simulation\label{S-Improvements}}

In order to ensure that the encouraging results of the \emph{sotl} methods
presented in the previous section were not an artifact of the simplicity of
the simulation, we made some improvements to make it more realistic. It was
good to have a simple environment at first, to understand better the basic
principles of the control methods. However, once this was achieved, more
complexity was introduced in the simulation to test the performance of the
methods more thoroughly. We developed thus a scenario similar to the one of 
\cite{FaietaHuberman1993}.

We introduced the traffic flow in four directions, alternating streets. This
is, arteries still consist of one lane, but the directions alternate:
southbound-northbound in vertical roads, and eastbound-westbound in
horizontal roads. Also, we introduced the possibility of having more cars
flowing in particular directions. This allows us to simulate peak hour
traffic, regulating the percentages of cars that will flow in vertical
roads, eastbound, or southbound roads\footnote{$\%horizontal=100-\%vertical$%
; $\%westbound=100-\%eastbound$; $\%northbound=100-\%southbound$}.

The most unrealistic feature of the first simulations was the torus. We
introduced an option to switch it off. Cars that exit the simulation are
removed from it. For creating new cars, gates are chosen with a probability
proportional to the car percentages at vertical, eastbound, and southbound
roads. At chosen gates (northbound, southbound, eastbound, or westbound), a
car will be created with a probability

\begin{equation}
P_{newc}=1-\frac{c}{c_{max}}  \label{eqPnewc}
\end{equation}

where $c$ is the current number of cars, and $c_{max}$ is the maximum number
of cars. Notice that without a torus, traffic jams are less probable, since
new cars cannot be fed into the system until there will be space. What
occurs is that for high densities, the actual number of cars can be less
than half of the number of $c_{max}$.

We also added a probability of turning at an intersection \thinspace $%
P_{turn}$. Therefore, when a car reaches an intersection, it will have a
probability \thinspace $P_{turn}$ of reducing its speed and turning in the
direction of the crossing street. This can cause cars to leave platoons,
which were more stable in the first series of experiments.

\section{Second Results\label{S-SecondResults}}

\begin{figure*}[tbp]
\begin{center}
\includegraphics[
natheight= 10.2359in, natwidth=6.6599in,
height=9.0000in, width=5.8558in]
{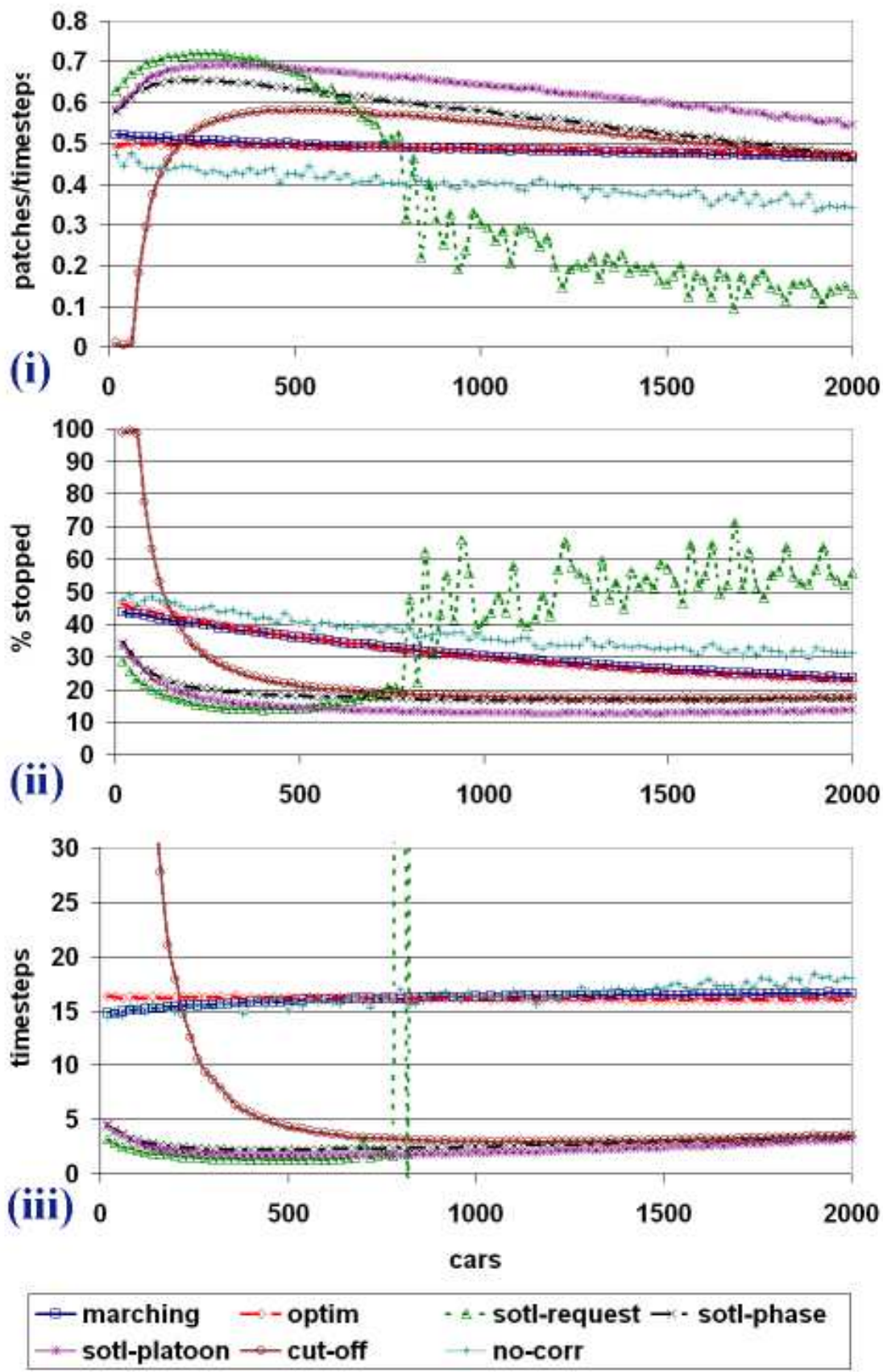}
\end{center}
\caption{Results in four directions, turning, and without torus, . (i)
Average speeds of cars. (ii) Percentage of stopped cars. (iii) Average
waiting times. (Color online).}
\label{4dresults}
\end{figure*}

\begin{figure*}[tbp]
\begin{center}
\includegraphics[
natheight=  4.1874in, natwidth=6.7136in, 
height=3.6523in, width=5.8558in]
{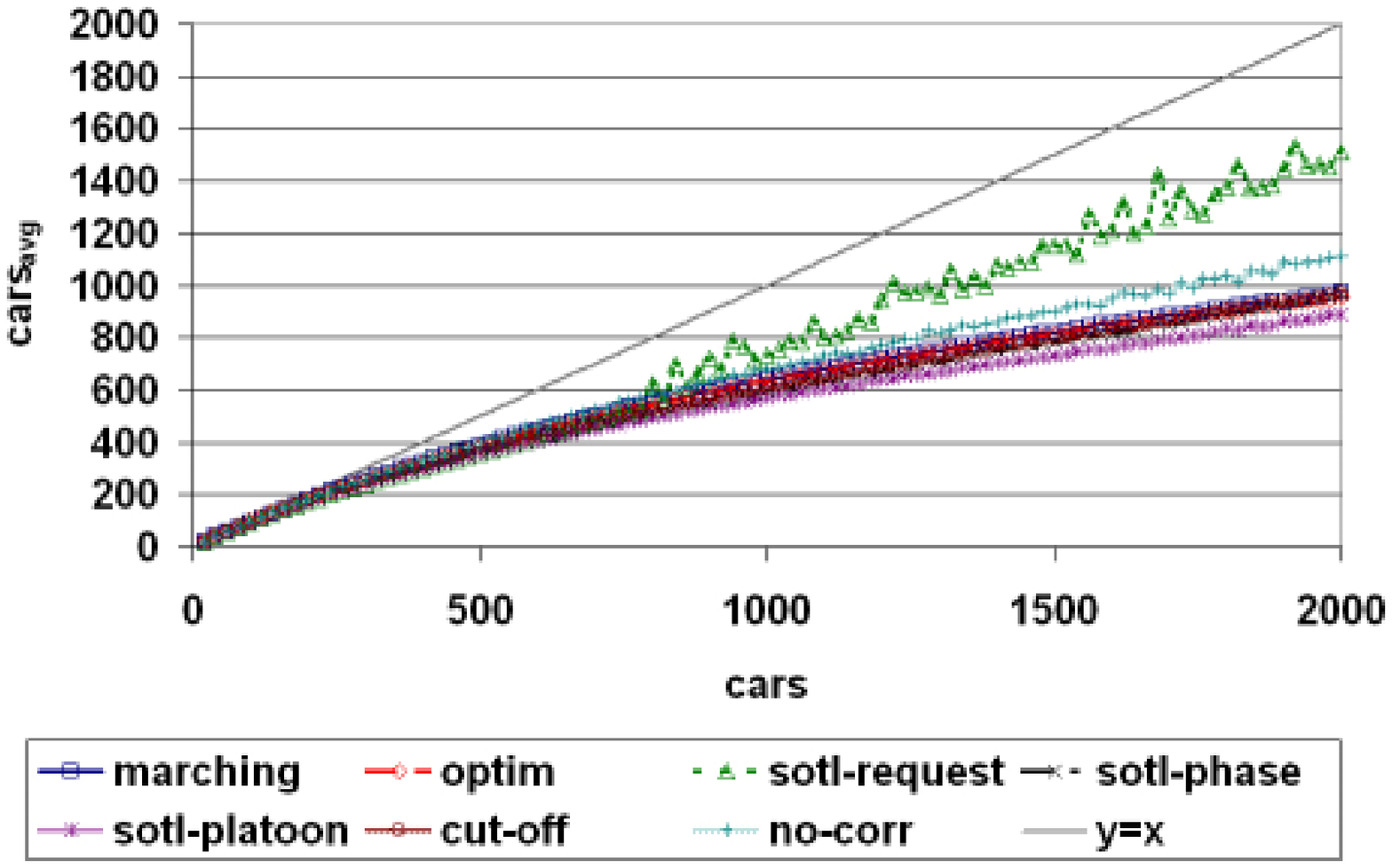}
\end{center}
\caption{Comparison of initial and average number of cars for different
methods without torus. (Color online).}
\label{carsiniavg}
\end{figure*}

We performed similar sets of experiments as the ones presented above. We did
runs of ten thousand time steps with random initial conditions in a grid of
10x10 arteries of $r=80$, with $p=83$, $\theta =41$, $\varphi _{\min }=20$, $%
\omega =4$, $\mu =3$, and $\lambda =3$. The percentage of cars in horizontal
streets was the same as in vertical, but of those, sixty percent in vertical
roads were southbound (forty percent northbound) and seventy five percent in
horizontal streets were eastbound (twenty five percent westbound). We used $%
P_{turn}=0.1$. Since each street crosses ten other streets, on average each
car should turn more than once. Results of singe runs, increasing the number
of initial cars ($c_{max}$ in equation (\ref{eqPnewc})) from twenty to two
thousand in steps of twenty, can be appreciated in Figure \ref{4dresults}.
We should note that the average number of cars is reduced as the initial
density increases, since cars cannot enter the simulation until there is
space for them. This reduces considerably the probability of deadlocks. We
can see a plot comparing the initial and average number of cars for the
simulations shown in Figure \ref{carsiniavg}.

In general terms, the improvements of the simulation did not alter much the
first results. \emph{Marching} and \emph{optim} are poor for low traffic
densities, but degrade slowly as the density increases. There are almost no
deadlocks because with high densities inserted in the simulation more cars
exit the simulation than those which enter. If this was a real city, there
would be queues waiting to enter the city, which the statistics of our
simulations do not consider.

\emph{Sotl-request} performs the best for low traffic densities, but worst
for high densities, even worse than \emph{no-corr}. This is because, as in
the first results, dense platoons force the traffic lights into a constant
switching, which reduces the performance.

The method \emph{sotl-phase} avoids this problem with the restriction set by 
$\varphi _{\min }$. It still performs very good for low densities, and the
average speeds degrades slowly to a comparable performance with the
non-adaptive methods. However, the percentage of stopped cars and the
waiting times are much lower than the non-adaptive methods.

\emph{Sotl-platoon} manages to keep platoons together, which enables them to
leave faster the simulation. It gives on average 30\% (up to 40\%) more
average speed, half the stopped cars, and seven times less average waiting
times than non-responsive methods. Therefore, this method performs the best
overall. It can adapt to different traffic densities, minimizing the
conflicts between cars. It is not possible to achieve almost perfect
performance, as it did for medium densities with a torus, since cars enter
the simulation randomly. Still, this method is the one that manages to adapt
as quickly as possible to the incoming traffic, organizing effectively
vehicles into platoons that leave quickly the simulation, even when single
vehicles might break apart from them (due to $P_{turn}>0$).

The \emph{cut-off} method again performs badly for very low densities.
Still, afterwards it performs better than the non-adaptive methods, but not
as good as \emph{sotl-phase} or \emph{sotl-platoon}.

Again, \emph{no-corr} shows that all methods give an improvement over random
phase assignment, except for \emph{sotl-request} at high densities, where
the method clearly breaks down.

The average number of cars, shown in Figure \ref{carsiniavg}, can be taken
as an indirect measure of the methods' performance: the faster the cars are
able to leave the simulation, there will be less cars in it, thus more
efficient traffic flow. We can observe an inverse correlation between the
average number of cars and the average speeds. If the traffic lights can
"get rid" of the incoming traffic as quickly as possible, it means that they
are successfully mediating the conflicts between vehicles.

The phenomenon of full synchronization is destroyed if there is no torus, or
if $P_{turn}>0$. However, it is still achieved when the cars flow in four
directions, or when the number of horizontal arteries is different from the
number of vertical arteries. It is easier to reach if there are less
arteries in the simulation. Also, if the length of horizontal and vertical
arteries differs, i.e. $r_{x}\neq r_{y}$, full synchronization is more
difficult to obtain, since the periods of the platoons passing on the same
traffic light depend on the length of the arteries. If these are
proportional, e.g. $r_{x}=2r_{y}$, full synchronization can be achieved.
Nevertheless, the \emph{sotl-phase} and \emph{sotl-platoon} methods achieve
very good performances under any of these conditions.

\section{Discussion\label{S-Discussion}}

In the series of experiments we performed, we could clearly see that \emph{\
sotl} strategies are more efficient than traditional control methods. This
is mainly because they are "\emph{sensitive}" and \emph{adaptive} to the
changes in traffic. Therefore, they can cope better with variable traffic
densities, noise, and unpredicted situations. Based on our results, we can
say the following:

\begin{itemize}
\item The formation of platoons can be seen as a reduction of \emph{variety} 
\cite[Ch. 11]{Ashby1956}. It is much easier to regulate ten groups of ten
cars than hundred cars independently\footnote{%
This could be seen as "functional" \emph{modularity \cite[pp. 188-195]%
{Simon1996}}}. Platoons make the traffic problem simpler. Oscillations in
traffic will be reduced if cars interact as groups. We can also see this as
a reduction of \emph{entropy}: if cars are homogeneously spread on the
street grid, at a particular moment there is the same probability of finding
a car on a particular block. This is a state of maximum entropy. However, if
there are platoons, there will be many blocks without any car, and few ones
with several. This allows a more efficient distribution of resources, namely
free space at intersections\footnote{%
The formation of platoons has already been proposed for freeways, with good
results (e.g. \cite{PATH1991})}. It is interesting to note that the \emph{%
sotl} methods do not \emph{force} vehicles into platoons, but \emph{induce}
them. This gives the system flexibility to adapt.

\item We can say that the \emph{sotl} methods try to "get rid" of cars as
fast and just as possible. This is because they give more importance to cars
waiting for more time compared to recently arrived ones, and also to larger
groups of cars. This successfully minimizes the number of cars waiting at a
red light and the time they will wait. The result is an increase in the
average speeds. Also, the prompt "dissipation" of cars from intersections
will prevent the formation of long queues, which can lead to traffic jams.

\item Since cars share a common resource --- space --- they are in
competition for that resource. Self-organizing traffic lights are \emph{%
synergetic\ \cite{Haken1981}}, trying to \emph{mediate} conflicts between
cars. The formation of platoons minimizes friction between cars because they
leave free space around them. If cars are distributed in a homogeneous way
in a city, the probability of conflict is increased.

\item There is no direct communication among the self-organizing traffic
lights. However, they "exploit" cars to transmit \emph{stigmergically}
information\footnote{%
For an introduction to stygmergy, see \cite{TheraulazBonabeau1999}}, in a
way similar to social insects exploiting their environment to coordinate.
For traffic lights, car densities form \emph{their} environment. Traffic
lights respond to those densities. But cars also respond to the traffic
light states. We could say that traffic lights and cars "co-control" each
other, since cars switch traffic lights to green, and red traffic lights
stop the cars.
\end{itemize}

\subsection{Adaptation or Optimization?}

Optimization methods are very effective for problems where the domain is
fairly static. This enables the possibility of searching in a defined space.
But in problems where the domain changes constantly, such as traffic, an
adaptive method should be used, to cope with these changes and constantly
approach solutions in an \emph{active} way.

The problem of traffic lights is such that cars and traffic lights face
different situations constantly, since they affect each other in their
dynamics (traffic lights affect cars, cars affect cars. With \emph{sotl}
methods also cars affect traffic lights and traffic lights affect other
traffic lights stigmergically via cars). If the situation is unknown or
unpredictable, it is better to use an adaptive, self-organizing strategy for
traffic lights, since it is not computationally feasible to predict the
system behaviour\footnote{%
This is because there is a high sensitivity to initial conditions in
traffic, i.e. chaos: if a car does not behave "as expected" by a
non-adaptive control system, this can lead the state of the traffic far from
the trajectory expected by the system}.

We can see an analogy with teaching: a teacher can tell exactly a student
what to do (as an optimizer can tell a traffic light what to do). But this
limits the student to the knowledge of the teacher. The teacher should allow
space for innovation if some creativity is to be expected. In the same way,
a designer can allow traffic lights to decide for themselves what to do in
their current context. Stretching the metaphor, we could say that the
self-organizing traffic lights are "gifted with creativity", in the sense
that they find solutions to the traffic problem by themselves, without the
need of the designer of even understanding the solution. On the other hand,
non-adaptive methods are "blind" to the changes in their environment, which
can lead to a failure of their rigid solution.

We can deduce that methods that are based on phase cycles, and even adaptive
cyclic systems \cite{SCATS1979,SCOOT1981}, will not be able to adapt as
responsively as methods that are adaptive and non-cyclic, since they are not
bounded by fixed durations of green lights \cite{PorcheLafortune1998}.
Therefore, it seems that optimizing phases of traffic lights is not the best
option, due to the unpredictable nature of traffic.

All traffic lights can be seen as a \emph{mediator} \cite{Heylighen2004}\
among cars. However, static methods do not take into account the current
state of vehicles. They are more "autocratic". On the other hand, adaptive
methods are regulated by the traffic flow itself. Traffic controls itself,
mediated by "democratic" adaptive traffic lights.

\subsection{Practicalities}

There are many parallel approaches trying to improve traffic. We do not
doubt that there are many interesting proposals that could improve traffic,
e.g. to calculate in real time trajectories of all cars in a city depending
on their destination via GPS. However, there are the feasibility and
economic aspects to take into account. Two positive points in favour of the
self-organizing methods is that it would be very easy and cheap to implement
them. There are already sensors on the market which could be deployed to
regulate traffic lights in a way similar to \emph{sotl-phase}. Sensors
implementing the \emph{sotl-platoon} method would not be too difficult to
deploy. Secondly, there is no need of a central computer, expensive
communication systems, or constant management and maintenance. The methods
are robust, so they can resist incrementally the failure of intersections.

Self-organizing traffic lights would also improve incoming traffic to
traffic light districts, e.g. from freeways, since they adapt actively to
the changing traffic flows. They can sense when more cars are coming from a
certain direction, and regulate the traffic equitatively.

Pedestrians could be included by in a self-organizing scheme considering
them as cars approaching a red light. For example, a button could be used as
the ones used commonly to inform the intersection, and this would contribute
to the count $\kappa _{i}$.

Vehicle priority could be also implemented, by simply including weights $%
w_{j}$ associated to vehicles, so that the count $\kappa _{i}$ of each
intersection would reach the threshold $\theta $ counting $w_{j}c\ast ts$.
However, this would require a more sophisticated sensing mechanism, although
available with current technology for priority vehicle detection. Still,
this would provide an adaptive solution for vehicle priority, which in some
cities (e.g. London) can cause chaos in the rest of the traffic lights
network, since lights are kicked off phase.

We should also note that traffic lights are not the best solution for all
traffic situations. For example, roundabouts \cite{FouladvandEtAl2004b}\ are
more effective in low speed, low density neighbourhoods.

\subsection{Unattended Issues}

The only way of being sure that a self-organizing traffic light system would
improve traffic is to implement it and find out. Still, the present results
are encouraging to test our methods in more realistic situations.

A future direction worth exploring would be a systematic exploration of the
parameters $\theta $, $p$, and $\varphi _{\min }$ values for different
densities. A meta-adaptive method for regulating these parameters depending
on the traffic densities would be desirable, but preliminary results have
been discouraging. In real situations this could be easier, because the
efficiency of different values can be tested experimentally for specified
traffic densities. Therefore, if a certain density is detected, proper
parameter values could be used. More realistic situations should be also
added to our simulations, such as multiple-street intersections,
multiple-lane streets, lane changing, different driving behaviours, and non
homogeneous streets. It would be also interesting to compare our methods
with others, e.g. \cite{SCATS1979,SCOOT1981}, but many of these are not
public, or very complicated to implement in reasonable time. Reinforcement
learning methods \cite{WieringEtAl2004} will adapt to a particular flow
density. However, in real traffic densities change constantly and unevenly.
We should compare the speed of adaptation of these methods with the proposed
self-organizing ones, but intuition tells us that learning methods will be
effective only for a particular fixed traffic density. We would also like to
compare our methods with other distributed adaptive \emph{cyclic} methods,
e.g. \cite{FaietaHuberman1993,Ohira1997} (\emph{sotl} and \emph{cut-off} are
non-cyclic), to test if indeed phase cycles reduce the adaptability of
traffic lights.

Another direction worth exploring would be to devise methods similar to the
ones presented that \emph{promote} "optimal" sizes of platoons for different
situations. We would need to explore as well which sizes of platoon yield
less interference for different scenarios.

\section{Conclusions\label{S-Conclusions}}

We have presented three self-organizing methods for traffic light control
which outperform traditional methods due to the fact that they are "aware"
of changes in their environment, and therefore are able to adapt to new
situations. The methods are very simple: they give preference to cars that
have been waiting longer, and to larger groups of cars. Still, they achieve
self-organization by the probabilistic formation of car platoons. In turn,
platoons affect the behaviour of traffic lights, prompting them to turn
green even before they have reached an intersection. Traffic lights
coordinate stigmergically via platoons, and they minimize waiting times and
maximize average speeds of cars. Under simplified circumstances, two methods
can achieve robust full synchronization, in which cars do not stop at all.

From the presented results and the ones available in the literature \cite%
{PorcheLafortune1998}, we can see that the future lies in schemes that are
distributed, non-cyclic, and self-organizing. In the far future, when
autonomous driving becomes a reality, new methods could even make traffic
lights obsolete \cite{Gershenson1998b,DresnerStone2004}, but for the time
being, there is much to explore in traffic light research.

There are several directions in which our models could be improved, which at
the present stage might be oversimplifying. However, the current results are
very promising and encourage us to test self-organizing methods in real
traffic environments.

\begin{acknowledgments}

Ricardo Barbosa, Vasileios Basios, Pamela Crenshaw, Kurt Dresner, Francis
Heylighen, Bernardo Huberman, Stuart Kauffman, Tom Lenaerts, Mike McGurrin,
Kai Nagel, Marko Rodriguez, Andreas Schadschneider, Seth Tisue, and Bart de
Vylder provided useful comments and assistance in the development of this
manuscript. This research was partially supported by the Consejo Nacional de
Ciencia y Teconolg\'{\i}a (CONACyT) of Mexico.

\end{acknowledgments}

\bibliographystyle{unsrt}
\bibliography{carlos,sos,traffic}

\end{document}